\documentclass[prd,showpacs,superscriptaddress,preprintnumbers,twocolumn,amsmath,nofootinbib]{revtex4}
 \usepackage[dvips,final]{graphicx}
  \usepackage{amssymb,amsmath,epsfig,bm,pifont}
   \graphicspath{{./figs/}}
\usepackage{color}

\usepackage{relsize}
\newcommand{\babar}{{\mbox{\slshape B\kern-0.1em{\smaller A}\kern-0.1em
            B\kern-0.1em{\smaller A\kern-0.2em R}}}
\def\MSbar{\relax\ifmmode\overline                        
            {\rm MS}\else{$\overline{\rm MS}${ }}\fi}     
           }                                              
\newcommand{\Ds}{\displaystyle}                           
\def\MSbar{\relax\ifmmode\overline                        
            {\rm MS}\else{$\overline{\rm MS}${ }}\fi}     
\def\1{\hbox{{1}\kern-.25em\hbox{l}}}

\begin{document}
\thispagestyle{empty}
 \date{\today}
  \preprint{\hbox{RUB-TPII-01/2016}}
\title{Systematic estimation of theoretical uncertainties in the
       calculation of the pion-photon transition form factor using
       light-cone sum rules}

\author{S.~V.~Mikhailov}
\email{mikhs@theor.jinr.ru}
\affiliation{Bogoliubov Laboratory of Theoretical Physics, JINR,
                141980 Dubna, Russia\\}

\author{A.~V.~Pimikov}
\email{pimikov@theor.jinr.ru}
\affiliation{Bogoliubov Laboratory of Theoretical Physics, JINR,
                141980 Dubna, Russia\\}
\affiliation{Institute of Modern Physics, Chinese Academy of Sciences,
                Lanzhou, 730000, P. R. China\\}

\author{N.~G.~Stefanis}
\email{stefanis@tp2.ruhr-uni-bochum.de}
\affiliation{Institut f\"{u}r Theoretische Physik II,
                Ruhr-Universit\"{a}t Bochum,
                D-44780 Bochum, Germany\\}

\begin{abstract}
We consider the calculation of the pion-photon transition form factor
$F^{\gamma^*\gamma\pi^0}(Q^2)$ within light-cone sum rules
focusing attention to the low-mid region of momenta.
The central aim is to estimate the theoretical uncertainties which
originate from a wide variety of sources related to
(i) the relevance of next-to-next-to-leading order
radiative corrections
(ii) the influence of the twist-four and the twist-six term
(iii) the sensitivity of the results on auxiliary parameters,
like the Borel scale $M^2$,
(iv) the role of the phenomenological description of resonances,
and
(v) the significance of a small but finite virtuality of the quasireal
photon.
Predictions for $F^{\gamma^*\gamma\pi^0}(Q^2)$ are presented which
include all these uncertainties and found to comply within the margin
of experimental error with the existing data in the $Q^2$ range
between 1 and 5 GeV$^2$, thus justifying the reliability of the applied
calculational scheme.
This provides a solid basis for confronting theoretical
predictions with forthcoming data bearing small statistical errors.
\end{abstract}

\maketitle

\section{Introduction}
\label{sec:intro}
During the last years, several experimental groups have reported data
on the pion-photon transition form factor (TFF).
Typically, these B factory experiments are single-tag
$\gamma^{*}(q_1)\gamma(q_2)\rightarrow \pi^0(P)$
measurements in which one of the two photons has a very small
virtuality $q^2_2\to 0$, inherited by the untagged electron, while the
other photon is highly off shell.
Therefore, the TFF measured in such an experimental setup is a function
of one --- the large $q_{1}^{2}=-Q^2$ --- photon virtuality,
$F_{\gamma\pi}(Q^2)$.
The recent theoretical interest focused primarily on the BABAR
experiment (2009) \cite{BaBar09} because of two reasons.
First, because it extended the range of data to quite high $Q^2$ values
of the order of $40$~GeV$^2$ and, second, because just these high-$Q^2$
data were found to increase with the momentum $Q^2$ --- an unexpected
result within the collinear factorization scheme of quantum
chromodynamics (QCD) \cite{Efremov:1978rn,BL80}.
The subsequent Belle experiment (2012) \cite{Belle12} covered the same
domain of momenta with similar precision, but did not confirm the
rising trend of the scaled $\pi\gamma$ TFF at high $Q^2$, with most
data points being in agreement with the hard-scattering limit of QCD.

Several theoretical groups have attempted to provide explanations for
the auxetic\footnote{The term auxetic was introduced and explained in
\cite{SBMP12}.
In the following it is used to describe the deviation from the
hard-scattering limit of QCD following from collinear factorization.}
behavior of the high-precision BABAR data presuming that these are also
accurate, i.e., true values and not the result of a false measurement.
These efforts range from approaches with the sole aim to provide
after-the-fact rationalizations of such an anomalous increase of the
scaled form factor
\cite{Dorokhov:2009dg,Radyushkin:2009zg,Polyakov:2009je,Li:2009pr}
--- to name just a few ---
to analyses arguing that the auxetic behavior of the BABAR data above
$\sim 10$~GeV$^2$ is incompatible with QCD and cannot be reproduced by
predictions obtained herewith see, for example,
\cite{MS09,BMPS11,BMPS12,SBMP12,Brodsky:2011yv,Brodsky:2011xx,Raya:2015gva}.
Under this particular perspective, the high-$Q^2$ BABAR data
are --- in the statistical sense --- precise but not accurate because
they fail to cluster around the ultraviolet (UV) limit,
$
 Q^2F^{\gamma^{*}\gamma\pi^{0}}(Q^2 \to \infty, 0)
=
 \sqrt{2}f_\pi$~GeV, which is an exact result of QCD
\cite{BL80,Brodsky:1981rp}.
Still other theorists \cite{ABOP10,ABOP12} argue that a best-fit to
\textit{all} high-$Q^2$ data (Belle and BABAR), being somewhere in
between (see \cite{BMPS12} for a classification scheme of theoretical
predictions), would only show a moderate increase of the scaled TFF at
currently accessible momenta so that this enhancement could still be
accommodated within the standard framework of QCD based on collinear
factorization without the need to invoke unconventional nonperturbative
mechanisms.
This treatment, they say, is justifiable, given that the relative
deviation between the Belle and the BABAR data fits does not exceed
$1.5\sigma-2\sigma$ \cite{Belle12}.
Moreover, it is not a priori known at which $Q^2$ values the TFF should
reach the asymptotic limit either from below or from above.
The issue around the incongruent trends of the high-$Q^2$ measurements
may be resolved after 2018 when the BelleII experiment at the SuperKEKB
collider in Japan will start collecting high-precision data on
two-photon physics, see, e.g., \cite{Wang:2015hdf}, so that the correct
behavior of the TFF at large momenta
$Q^2F^{\gamma^{*}\gamma\pi^{0}}(Q^2\gg 1 \text{GeV}^2, 0)$
can be estimated more rigorously, eventually reducing the range of
multi-layered theoretical predictions to a single reliable
curve within a comparably small margin of systematic theoretical
error \cite{BMPS12}.

Despite this debate at the high-end of the probed momentum values in
the measurement of the pion-photon TFF also the mid-low-$Q^2$ region
is of particular importance.
The reason is that the available data sets obtained
in the range $[1 - 5]$~GeV$^2$ nearby the modern normalization scale
$\mu_0=2$~GeV, used in lattice simulations and other calculations
\cite{Chang:2013pq,Stefanis:2014nla,Stefanis:2015qha},
have rather large errors so that they cannot be used to fine-tune
theoretical predictions in this domain.
This applies to the CELLO \cite{CELLO91} data and partially also to
the CLEO \cite{CLEO98} data.
The situation is expected to improve significantly when the data
of the BESIII Collaboration, taken with the single-tag technique at
$\sqrt{s}=3.770$~GeV with the BESIII detector at the BEPCII collider,
will become available.
The process under study is
$e^+ e^- \to e^+ e^- \text{hadron(s)}$,
where either the electron or positron in the final state is detected.
However, for the time being, only simulated data in the range
$Q^2\in [0.5-3]$~GeV$^2$
have been publicized which mainly serve to demonstrate the small
size of the experimental errors in the event analysis
\cite{Denig:2014mma}.
Assuming as a pretext that the BESIII Collaboration will indeed provide
real data with very small statistical errors in the spacelike region
$Q^2 \lesssim 4$~GeV$^2$,
we may attempt to quantify how the existence of such data might be used
to confront in more detail the theoretical systematic uncertainties,
pertinent to the employed calculational method in this momentum
regime.
Such dedicated theoretical investigations have been carried out before
within particular approaches.
These include soft QCD modeling based on a set of Dyson-Schwinger
equations (DSE) truncated to the ladder-rainbow level
\cite{Maris:2002mz}, or employ ideas related to the vector-meson
dominance and the Pad\'e approximation \cite{Masjuan:2012wy}.
In a more recent work, the pion TFF was calculated by means of a
dispersive approach in terms of the most important intermediate states
\cite{Hoferichter:2014vra}.
The small to medium $Q^2$ region was also addressed within
AdS/QCD using a holographic confining model in terms of an effective
interaction in light-front time \cite{Brodsky:2011xx}.
In the context of the light-cone sum-rule (LCSR) method such analysis
has not yet been carried out and is part of the present investigation.

Several challenging questions arise:
(i) How significant is the inclusion of higher twists, e.g.,
twist-four and twist six, at scales around 1-2~GeV$^2$ relative to the
leading twist-two term?
(ii) Are radiative corrections at the next-to-next-to-leading order
(NNLO) level relevant at such low momentum scales?
(iii) How reliable are light-cone sum rules for the calculation of
$F_{\gamma\pi}(Q^2)$ in the $Q^2 \sim 1 $~GeV$^2$ region?
(iv) How strong is the influence of the finite virtuality of the
quasireal photon at such scales?
This work seeks quantitative answers to these questions.

The plan of the paper is the following.
In the next section, we will examine the pion-photon TFF making use of
QCD factorization to be followed in Sec.\ \ref{sec:lcsr} by its
formulation in the framework of LCSRs.
To incorporate the nonperturbative input of the pion bound state
of twist two, the BMS\footnote{The acronym BMS is a reference to the
authors of Ref.\ \cite{BMS01}.} distribution amplitudes (DA)s and the 
platykurtic DA \cite{Stefanis:2014nla,Stefanis:2015qha} will be used.
The main radiative corrections (up to the NNLO level) and the key
higher-twist contributions (twist-four and twist six) to the TFF will
be considered in Sec.\ \ref{sec:LCSR-uncertainties}.
Section \ref{sec:LCSR-predictions} is devoted to the comparison of the
obtained predictions with the low-$Q^2$ data, the particular emphasis
being placed on the new elements of our upgraded theoretical framework
and the estimation of the most crucial systematic uncertainties.
A summary of our findings and our conclusions will be given in Sec.\
\ref{sec:concl}.
Some important technical ingredients of the approach are provided in
two appendices.

\section{Pion-photon transition form factor using QCD factorization}
\label{sec:fact-scheme}
Let us begin our analysis by considering the process
$\gamma^*(q_{1}^{2})\gamma(q_{2}^{2})\to \pi^0$,
with $q_{1}^{2}=-Q^2$ for the far-off shell photon and
$q_{2}^{2}=-q^2 \gtrsim 0$ for the quasireal photon,
described by the pion-photon transition form factor
\begin{eqnarray}
&& \int\! d^{4}z\,e^{-iq_{1}\cdot z}
  \langle
         \pi^0 (P)| T\{j_\mu(z) j_\nu(0)\}| 0
  \rangle
=
  i\epsilon_{\mu\nu\alpha\beta}
  q_{1}^{\alpha} q_{2}^{\beta}
\nonumber \\
&&  ~~~~~~~~~~~~~~ \times ~
  F^{\gamma^{*}\gamma^{*}\pi^0}(Q^2,q^2)\ ,
\label{eq:matrix-element}
\end{eqnarray}
where $j_\mu$ is the quark electromagnetic current.
Employing perturbative QCD (pQCD) in connection with collinear
factorization, the leading-twist two TFF for two highly off-shell
photons,
$F^{\gamma^{*}\gamma^{*}\pi^0}(Q^2,q^2;\mu_{\rm F}^2)$,
can be cast in convolution form at the factorization scale
$\mu_{\rm F}^2$ to read
\begin{eqnarray}
  F^{\gamma^{*}\gamma^{*}\pi^0}\!\!\left(Q^2,q^2;\mu_{\rm F}^2\right)
\! = \!
  T\left(Q^2,q^2;\mu_{\rm F}^2;x\right)
\!\otimes\!
  \varphi_{\pi}^{(2)}\!\left(x,\mu_{\rm F}^2\right) \, .
\label{eq:fact-TFF}
\end{eqnarray}
Here, $\otimes \equiv \int_{0}^{1} dx$
and
$\varphi_{\pi}^{(2)}\left(x,\mu_{\rm F}^2\right)$
denotes the pion distribution amplitude of leading twist two.
It describes the partition of longitudinal-momentum fractions between
the valence quark ($x_q=x=(k^0+k^3)/(P^0+P^3)=k^+/P^+$) and antiquark
($x_{\bar{q}}=1-x\equiv \bar{x}$)
at the scale $\mu_{\rm F}$.
The hard-scattering amplitude $T$ can be expressed as a power-series
expansion in the running strong coupling
$a_s\equiv \alpha_{s}(\mu_{\rm R}^2)/4\pi$
to obtain
\begin{equation}
  T\left(Q^2,q^2;\mu_{\rm F}^2;x\right)
=
  T_\text{LO} + a_s~T_\text{ NLO} + a_{s}^2~T_\text{NNLO} + \ldots \, ,
\label{eq:hard-scat-ampl}
\end{equation}
where we have adopted the so-called default scale setting in which
the renormalization scale $\mu_{\rm R}$ is set equal to the
factorization scale $\mu_{\rm F}$, i.e.,
$a_s\left(\mu_{\rm R}^{2}\right)=a_s\left(\mu_{\rm F}^{2}\right)$.
We have also used for convenience the following abbreviations:
LO --- leading order,
NLO --- next-to-leading order, and
NNLO --- next-to-next-to-leading order.
The corresponding contributions will be labeled by the superscripts
$(0)$, $(1)$, and $(2)$, respectively; they indicate the power of
the strong coupling $a_s$.

The various terms in (\ref{eq:hard-scat-ampl}) are given by the
following expressions
\begin{subequations}
\label{eq:T}
\begin{eqnarray}
  T_{\rm LO}
& = &
  T_0,
\\
  T_{\rm NLO}
& = & C_{\rm F}~
  T_0 \otimes \left[\mathcal{T}^{(1)}+
                      L~  V_{+}^{(0)}
              \right],
\label{eq:NLO}
\\
  T_{\rm NNLO}
& = &C_{\rm F}~
  T_0 \otimes \left[
                     {\cal T}^{(2)} + L~ V_{+}^{(1)}/C_{\rm F}
                     - L~ \beta_{0}  {\cal T}^{(1)} \right.  \nonumber \\
                     && - \left.
                     \frac{L^2}{2}~ \beta_{0} V_{+}^{(0)}
                     + \frac{L^2}{2}~ C_{\rm F} V_{+}^{(0)}\otimes V_{+}^{(0)}\right.
\nonumber \\
                     && + \left.
                     L~ C_{\rm F} {\cal T}^{(1)}\otimes V_{+}^{(0)}
              \right] \, ,
\label{eq:hard-scat-series}
\end{eqnarray}
\end{subequations}
in which we have introduced the convenient abbreviation
$
 L
\equiv
 \ln\left[\left(Q^2y+q^2\bar{y}\right)/\mu^2_\text{F}\right]
$, see \cite{Melic:2002ij}.
Each term of the hard-scattering amplitude comprises several
contributions originating from different sources.
A common factor is the lowest-order (Born) term, viz.,
$T_0(Q^2, q^2;y)$,
while
$\mathcal{T}^{(1)}(y,x)$ and $\mathcal{T}^{(2)}(y,x)$
represent the coefficient functions of the considered partonic
subprocess.
The NLO contribution $T_{\rm NLO}$ in Eq.\ (\ref{eq:NLO}) is
completely known \cite{BMS02,Melic:2002ij}.
On the other hand, the NNLO correction $T_{\rm NNLO}$
(see Eq.\ (\ref{eq:hard-scat-series}))
contains the quantities $V^{(0)}(y,x)$ and $V^{(1)}(y,x)$,
which denote, respectively, the one- and two-loop kernels of the
Efremov-Radyushkin--Brodsky-Lepage (ERBL) \cite{Efremov:1978rn,BL80}
evolution equation.
Their structures are displayed explicitly in Appendix
\ref{sec:nlo-evol-kern}, using for $V_{+}^{(1)}$ a new more compact
representation, derived in this work, which is given by
Eq.\ (\ref{eq:V1-structure}).
Note that $\beta_0$ is the first coefficient of the QCD
$\beta$-function displayed in Eq.\ (\ref{eq: beta0}).
Furthermore, we isolate the important term $T_\beta$ \cite{MS09},
which accumulates all terms proportional to $\beta_0$ on the RHS
of Eq.\ (\ref{eq:hard-scat-series}), to obtain
\begin{eqnarray}
\beta_0 T_\beta
\!& = &\!
\beta_0 \left[\mathcal{T}_\beta^{(2)} + L\left( V_{\beta +}^{(1)} -
                                      \mathcal{T}^{(1)} \right)
                                   - \frac{L^2}{2} V^{(0)}_{+}
 \right] ,
\label{eq:hard-scat-nnlo-beta}
\end{eqnarray}
where on the RHS we used the known decompositions of the kernel
\cite{MR86ev,MS09} and the NNLO coefficient function determined
in \cite{Melic:2002ij}:
\begin{subequations}
\label{eq:mainTbeta}
\begin{eqnarray}
  V^{(1)}/C_{\rm F}
&=&
  \beta_0 V_{\beta}^{(1)} + \Delta V^{(1)}, \label{eq:V-NLO} \\
  \mathcal{T}^{(2)}
&=&
  \beta_0 \mathcal{T}_\beta^{(2)}+\mathcal{T}^{(2)}_{c} \, .
\label{eq:dv-term}
\end{eqnarray}
\end{subequations}
With the help of these expressions, $T_{\rm NNLO}$ can be recast in the
more compact form
\begin{eqnarray}
  T_{\rm NNLO}
= C_{\rm F} T_0 \otimes \!
    \left[\beta_0 T_\beta
  + T_{\Delta V}
  + T_L+  \mathcal{T}^{(2)}_c \right]\, ,
\label{eq:NNLO-beta}
\end{eqnarray}
where
\begin{subequations}
 \label{eq:T-elements}
\begin{eqnarray}
   T_{\Delta V}
& = &
 L \Delta V^{(1)}_+,
\label{eq:hard-scat.nnlo-dv} \\
  T_L
& = &
 C_{\rm F} L \left(
           \frac{L}{2} V_{+}^{(0)}\otimes V_{+}^{(0)}
+ \mathcal{T}^{(1)}\otimes V_{+}^{(0)}
             \right)
                     \, .
\label{eq:hard-scat-nnlo}
\end{eqnarray}
It is useful to express the elements in
Eqs.\ (\ref{eq:T}), (\ref{eq:mainTbeta}), (\ref{eq:T-elements})
in convolution form by employing the eigenfunctions $\psi_n(x)$ of the
LO ERBL evolution equation.
This leads to simpler expressions, e.g., (\ref{eq:hard-scat-nnlo})
becomes (arguments suppressed)
\begin{eqnarray}
  T_L \otimes \psi_{n}
& \! = \! &
  2C_{\rm F}L \, v(n)\left[
  L v(n) \psi_{n}
  + \mathcal{T}^{(1)}\otimes \psi_{n}
                   \right] \, ,
\end{eqnarray}
\end{subequations}
where the quantities $ v(n)$ denote the eigenvalues of the ERBL
evolution kernel
$V^{(0)}_+\otimes \psi_{n} = 2v(n)\psi_{n}$,
while the eigenfunctions $\psi_{n}$ can be expressed in terms of the
conformal basis of the Gegenbauer harmonics:
$\psi_{n}(x)=6x\bar{x}C_{n}^{(3/2)}(x-\bar{x})$.
This representation will be further used in Sec.\ \ref{sec:hi-rad-cor}
in connection with the construction of the spectral density.

At the NNLO level we note the following.
The main contribution
$\beta_0\mathcal{T}_\beta^{(2)}$
to
$T_{\rm NNLO}$
in (\ref{eq:NNLO-beta}) has been calculated in \cite{Melic:2002ij},
whereas the terms $T_{\Delta V}$ and $T_L$ in the form they enter the
corresponding contributions to the spectral density are derived here
and are presented in Appendix \ref{sec:spectr-dens}.
Finally, the term
$\mathcal{T}^{(2)}_c$
represents the still uncalculated part of $T_{\rm NNLO}$.

The physics of nonperturbative interactions in the TFF is included by
means of the leading-twist pion DA
$\varphi_{\pi}^{(2)}\left(x,\mu_{\rm F}^2\right)$
which is defined by the following gauge-invariant matrix element
\begin{eqnarray}
  \langle 0| \bar{q}(z) \gamma_\mu\gamma_5 [z,0] q(0)
           | \pi(P)
  \rangle|_{z^{2}=0}
&& \!\!\! \! \! =
  if_\pi P_\mu \int_{0}^{1} dx e^{i x (z\cdot P)}
\nonumber \\
&& \times \varphi_{\pi}^{(2)} \left(x,\mu^2\right) \, ,
\label{eq:pion-DA}
\end{eqnarray}
where the lightcone gauge $A^+=0$ is to be imposed so that
$[z,0]=1$, i.e., the gauge link reduces to the identity operator.

\section{Light-cone sum rules for the pion-photon TFF}
\label{sec:lcsr}
As mentioned in the Introduction, the existing experimental data
at low $Q^2$ values are not precise enough to allow for reliable
information extraction on the detailed behavior of the TFF in
terms of magnitude and slope.
This, however, would be extremely valuable given that theoretical
calculations are only approximations and one needs some quantitative
etalon to estimate more precisely their range of reliability that is
intimately related to various perturbative and nonperturbative
contributions with their own sources of uncertainties.
The publication of the BESIII data may change this situation
significantly.
Our particular aim in this paper is to work out the applicability
limits of our LCSR-based approach for the calculation of the
pion-photon TFF in the low-$Q^2$ regime in anticipation of this set of
data.
To this end, let us now consider the pion-photon TFF in more detail
within the method of LCSRs, which is based on the operator product
expansion on the lightcone and enables the systematic computation of
QCD radiative corrections and higher-twist contributions.

Within this approach, the form factor for the $\pi\to \gamma$
transition is described in terms of a dispersion integral
which employs the spectral density
\begin{equation}
  \bar{\rho}(Q^2,x)
=
  (Q^2+s) \rho^{\text{pert}}(Q^2,s) \, .
  \label{eq:rho-bar}
\end{equation}
The quantity $\rho^{\text{pert}}(Q^2,s)$ is given by
\begin{eqnarray}
 \label{eq:rho}
  \rho^{\text{pert}}(Q^{2},s)
& = &
  \frac{1}{\pi} {\rm Im}F^{\gamma^*\gamma^*\pi^0}_\text{pert}
  \left(Q^2,-s-i\varepsilon\right)
\end{eqnarray}
and can be calculated in fixed-order QCD perturbation theory.
Then, taking into account
that $s =\bar{x}Q^2/x$, the TFF assumes the following form
\begin{widetext}
\begin{eqnarray}
  Q^2 F^{\gamma^*\gamma\pi}\left(Q^2\right)
=
  \frac{\sqrt{2}}{3}f_\pi
  \left[
        \frac{Q^2}{m_{\rho}^2}
        \int_{x_{0}}^{1}
        \exp\left(
                  \frac{m_{\rho}^2-Q^2\bar{x}/x}{M^2}
            \right)
        \! \bar{\rho}(Q^2,x)
  \frac{dx}{x}
  + \! \int_{0}^{x_0} \bar{\rho}(Q^2,x)
        \frac{dx}{\bar{x}}
  \right]
\, .
\label{eq:LCSR-FQq}
\end{eqnarray}
\end{widetext}
The expression on the RHS of Eq.\ (\ref{eq:LCSR-FQq}) depends on
various parameters and is bounded in the region
$\Ds x_0 = Q^2/\left(Q^2+s_0\right)$.
In our present analysis, the Borel parameter $M^2$ is taken to vary
in the interval $[0.7-1.0]$~GeV$^2$ as in our previous works
\cite{BMPS11,BMPS12,SBMP12}.
But in order to estimate the uncertainty due to the variation of this
parameter, we also consider the larger value $M^2=1.5$~GeV$^2$ employed
in \cite{ABOP10,ABOP12}.
The duality interval in the vector channel is assumed to be
$s_0\simeq 1.5$~GeV$^2$,
whereas $m_\rho=0.77$~GeV \cite{Agashe:2014kda}, and
the pion decay constant has the value $f_\pi=132$~MeV.
Expression (\ref{eq:LCSR-FQq}) represents a sum rule which makes use
of a simple $\delta$-function \textit{ansatz} to model the
$\rho$-meson resonance.
In the real calculation carried out here, the $\rho$ and $ \omega$
resonances are taken
into account in terms of the Breit-Wigner (BW) form, as done before in
\cite{MS09}.

The leading-twist pion DA $\varphi_{\pi}^{(2)}$ is expanded in
terms of the eigenfunctions $\psi_{n}(x)$ to read
\begin{equation}
  \varphi_{\pi}^{(2)}(x,\mu^{2})
=\psi_{0}(x)
  + \sum_{n=2,4, \ldots}^{\infty} a_{n}(\mu^{2}) \psi_{n}(x)
\label{eq:gegen-exp}
\end{equation}
and satisfies the normalization condition
$
 \int_{0}^{1}dx\varphi_{\pi}^{(2)}(x, \mu^{2})
=
 1
$,
so that
$\psi_{0}(x)=\varphi_{\pi}^{\rm asy}=6x\bar{x}$
is the asymptotic (asy) DA.
The conformal coefficients $a_{n}(\mu^{2})$ encode the nonperturbative
information and are not calculable within pQCD.
In our analysis we will consider various model DAs for the pion
pertaining to different nonperturbative approaches from which these
coefficients are determined.
For the sake of definiteness, the numerical uncertainty estimation
procedure in our analysis will be based on the set of the BMS DAs,
determined in \cite{BMS01} using QCD sum rules with nonlocal
condensates (NLC)s.
This choice introduces some bias but it is not conflicting with
observations \cite{BMPS12} and does not lead to an underestimation of
the size and influence of the theoretical uncertainties.
Moreover, it should not be understood as the result of a priori
justification of these DAs.
The low-$Q^2$ data alone are not sufficient to draw definite
conclusions about the shape of the pion DA.

Turning our attention to the spectral density, we first note that each
contribution of definite twist (tw) to $\rho^{\text{pert}}$ in
Eq.\ (\ref{eq:rho}), can be obtained from the convolution of the
associated hard part with the corresponding pion DA of the same twist
\cite{Kho99} so that one gets
\begin{eqnarray}
  \rho^{\text{pert}}(Q^{2},s)
& = &
   \rho_{\text{tw-2}}
  +\rho_{\text{tw-4}}
  +\rho_{\text{tw-6}}
  +\ldots\, .
\label{eq:rho-twists}
\end{eqnarray}
We then express the twist-two part of $\bar{\rho}(Q^{2},x)$ as a sum
over the partial spectral densities $\bar{\rho}_{n}$ each related to
a particular harmonic $\psi_{n}$.
In this way, we obtain ($a_0=1$)
\begin{eqnarray}
  \bar{\rho}\left(Q^2,x\right)
& = &
  \sum_{n=0,2,4,\ldots}a_{n}\left(Q^2\right)
    \bar{\rho}_{n}\left(Q^2,x\right)\nonumber \\
&&
  + \bar{\rho}_\text{tw-4}\left(Q^2,x\right)
  + \bar{\rho}_\text{tw-6}\left(Q^2,x\right) \, ,
\label{eq:rho-bar-15}
\end{eqnarray}
where
\begin{eqnarray}
  \bar{\rho}_{n}\left(Q^2,x\right)
\!\!\!&\!\!=\!\! &\!\!\!
    \bar{\rho}_{n}^{(0)}(x)
\!\! +\!a_{s}\bar{\rho}_{n}^{(1)}(Q^2,x)
\!\!+\!a_{s}^{2}\bar{\rho}_{n}^{(2)}(Q^2,x)
    + \ldots, \nonumber \\
\bar{\rho}_{n}^{(0)}(x)&=& \psi_n(x);~\, a_s= a_{s}(Q^2) \, .
\label{eq:rho-n}
\end{eqnarray}
The various terms of the spectral density in (\ref{eq:rho-n}) are the
key computational ingredients in our dispersion-relation-based LCSR
analysis and are therefore given explicitly in Appendix
\ref{sec:spectr-dens}.

The second term in Eq.\ (\ref{eq:rho-twists}) is the twist-four
contribution to the spectral density which reads
\cite{Kho99}
\begin{equation}
  \bar{\rho}_{\text{tw-4}}(Q^2,x)
=
  \frac{\delta^2_\text{tw-4}(Q^2)}{Q^2}
  x\frac{d}{dx}\varphi^{(4)}(x)\, ,
\label{eq:rho-tw-4}
\end{equation}
with the twist-four coupling parameter being given by
$
 \delta^2_\text{tw-4}
\approx \!\!
 (1/2)\lambda^{2}_{q}
\!\!=\!\!
 (1/2)\left(0.4 \pm 0.05 \right)$~GeV$^2$
at $Q^2\!\! \approx \!\! 1$ GeV$^2$ \cite{BMS02},
where $\lambda^{2}_{q}$ denotes the average virtuality of vacuum quarks
\cite{MR86}.
The full twist-four pion DA --- which originates from the contributions
of the two- and three-particle twist-four DAs --- is approximated here
by its asymptotic form \cite{Kho99}
\begin{equation}
\varphi_{\pi}^{(4)}(x)= \frac{80}{3}x^2(1-x)^2 \, ,
\label{eq:tw-4-DA}
\end{equation}
see \cite{ABOP10} (Sec.\ 3~C there) for further discussion.

The twist-six term
$\bar{\rho}_{\text{tw-6}}(Q^{2},x)
=
 (Q^2+s)\rho_{\text{tw-6}}(Q^2,s)
$
in Eq.\ (\ref{eq:rho-twists})
was first computed in \cite{ABOP10} (Eq.\ (58) there) and is given by
the following expression
\begin{widetext}
\begin{eqnarray}
 \label{eq:tw-6}
    \bar{\rho}_\text{tw-6}(Q^{2}\!,x)
=
    8\pi \frac{C_F}{N_c}
    \frac{ \alpha_s\langle\bar{q} q\rangle^2}{f_\pi^2}\frac{x}{Q^4}
    \left[
        \!-\!
        \left[\frac{1}{1-x}\right]_+
        \!+\!\left(2\delta(\bar{x})-4 x\right)\!+\!
        \left(
         3x+2x\log{x}
        \!+\!
        2x\log{}\bar{x}
        \right)
    \right] \, ,
\end{eqnarray}
\end{widetext}
where
$\alpha_s=0.5$ and $
 \langle \bar{q} q\rangle^2
=
 \left(0.242 \pm 0.01 \right)^6$ GeV$^6$ \cite{Gelhausen:2013wia}.
We have independently verified and confirmed this expression, which
may be considered as an inverse power correction to the coefficient
function.
It is important to make a remark on the structure of
Eq.\ (\ref{eq:tw-6}) in conjunction with the diagrams in Fig.\ 4 in
\cite{ABOP10}:
The first term in the square brackets originates from diagram (a),
while the second one stems from diagram (b), and the third one
derives from diagrams (c) and (d).
To discuss the structure of the LCSR in Eq.\ (\ref{eq:LCSR-FQq}),
it is useful to do it in comparison with the pQCD factorization
formula, looking more closely and critically at the behavior of the
TFF in the low to intermediate $Q^2$ region, say, between
1 and 5~GeV$^2$.
As we will make more explicit below, the main effect in using the
LCSR instead of the pQCD expression is the possibility of a
successive inclusion into the TFF of the higher harmonics
$\psi_{n>0}(x)$ as $Q^2$ grows.
This effect can be revealed already at the level of the leading-order
approximation of both expressions.

To this end, consider the contribution of a given harmonic to the
expression in the square brackets in Eq.\ (\ref{eq:LCSR-FQq}) and
approximate the perturbative part of the spectral density by
$\bar{\rho}(Q^2,x) \to \bar{\rho}^{(0)}_n(x)=\psi_n(x)$,
cf.\ (\ref{eq:rho-n}).
This way, we obtain a physical correspondence between the LCSR on the
left below
\begin{widetext}
\begin{eqnarray}
\text{LCSR} \phantom{\frac{Q^2}{m_{\rho}^2}
     e^{\frac{m^2_{\rho}}{M^2}} \int_{x_{0}}^{1}
       \Ds e^{ -\frac{Q^2\bar{x}}{M^2 x}}
~\psi_{n}(x)
  \frac{dx}{x}
  + \! \int_{0}^{x_0} \psi_{n}(x)
        \frac{dx}{\bar{x}}} &\Leftrightarrow&\phantom{Q^2F_n^\text{pQCD}(Q^2)}
        \text{ pQCD}, \nonumber \\
Q^2F_n^\text{LCSR}(Q^2)=
\frac{Q^2}{m_{\rho}^2}
     e^{\frac{m^2_{\rho}}{M^2}} \int_{x_{0}}^{1}
        \Ds e^{ -\frac{Q^2\bar{x}}{M^2 x}}
        ~\psi_{n}(x)
  \frac{dx}{x}
  + \! \int_{0}^{x_0} \psi_{n}(x)
        \frac{dx}{\bar{x}}
& \Leftrightarrow&
Q^2F_n^\text{pQCD}(Q^2)=
\int_{0}^{1}\psi_{n}(x)
  \frac{dx}{\bar{x}}=3\, .
\label{eq:LCSR-pQCD}
\end{eqnarray}
\end{widetext}
and the lowest-order leading-twist contribution from pQCD, shown on the
right, which amounts to the inverse moment of
$\psi_n$ on account of
$6\int_{0}^{1} dx x C_{n}^{(3/2)}(x-\bar{x})=3$ for any $n$.

\begin{figure}[th]
 \includegraphics[width=0.5\textwidth]{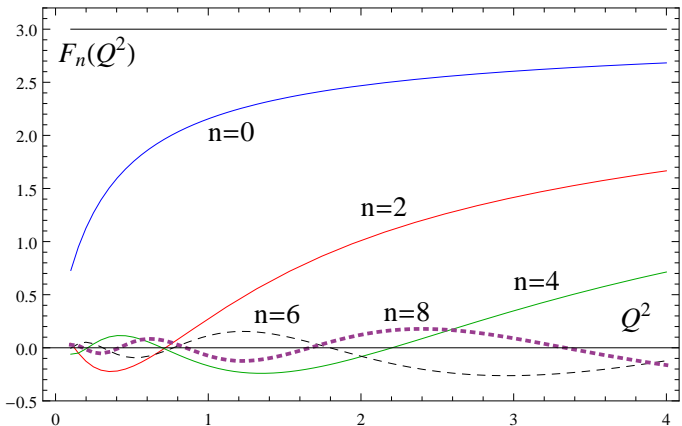} 
  \caption{
Contributions of $F^\text{LCSR}_n(Q^2)$ to
the TFF entering the left side of
(\ref{eq:LCSR-pQCD}) and originating from the successive inclusion of
Gegenbauer harmonics of increasing order $n$.
The harmonics with $n=$ 0, 2, 4, 6, 8 are shown explicitly using the
following designations from top to bottom:
$n=0$ --- upper solid blue line,
$n=2$ --- middle solid red line,
$n=4$ --- solid green line,
$n=6$ --- dashed black line,
$n=8$ --- dotted pink line.
All harmonics, except $\psi_0$, have a zero crossing
in the vicinity of $Q^2 \approx 0.8$~GeV$^2$.
The topmost solid horizontal line corresponds to the $F^\text{pQCD}_n$
result in (\ref{eq:LCSR-pQCD}) (right side).
\label{fig:knot}
}
\end{figure}

This correspondence can be completely vindicated by the following
observations:
\begin{itemize}
\item
For $Q^2 \gg s_0,~x_0=(1+s_0/Q^2)^{-1} \to 1$ and, employing the values
of $m_\rho$, $s_0$ and $M^2$ given farther above, the first term in the
LCSR on the left, which models the hadronic content of the quasireal
photon, becomes suppressed with $Q^2$.
Hence, the whole expression tends to the pQCD result, shown on the
right of (\ref{eq:LCSR-pQCD}), establishing also a \emph{mathematical}
correspondence between the LCSR and the pQCD expression.
In this latter result, all harmonic contributions of expansion
(\ref{eq:gegen-exp}) \textit{appear at once} -- see the horizontal
uppermost line in Fig.\ \ref{fig:knot}.

\item
In the opposite kinematic region $Q^2 \lesssim s_0,~x_0 \lesssim 1/2$,
both terms on the left are of the same order of magnitude and hence the
result differs strongly from that on the right, implying that LCSRs and
pQCD lead to different predictions for the TFF.
This is, mainly because higher twists, contributing via the first term
in the LCSR, are not accounted for in the pQCD expression.
This difference ensues from the treatment (in the LCSR) of the
quasireal photon by means of the vector-meson dominance model.
This model has been used to construct the phenomenological spectral
density and effectively takes into account long-distance gluon
interactions pertaining to this photon vertex \cite{Kho99}.
The key parameter to manage these long-distance (nonperturbative)
effects is $s_0$, the duality interval, and encompasses the masses of
the vector-meson family $m_{\rho},\ldots$ entering the first term on
the left in (\ref{eq:LCSR-pQCD}).
\end{itemize}

On the other hand, the Gegenbauer harmonics are included into the TFF
sequentially, i.e., term by term with increasing index $n$, in
correspondence with the growth of $Q^2$, giving rise to an oscillatory
behavior.
These zero crossings of the harmonics accumulate in the vicinity of
1~GeV$^2$ to build a knot (see Fig.\ \ref{fig:knot}.)
%
The contributions stemming from different harmonics vanish near the
first knot at $Q^2 \approx 0.8$~GeV$^2$, so that only the term with
$n=0$ survives which does not oscillate but grows uniformly.
This leads for momentum values $Q^2 \lesssim 1.4$~GeV$^2$ to the
dominance of the zero harmonic ($\psi_0$) contribution to the TFF.
At higher $Q^2$ values, the contributions stemming from higher
harmonics, beginning with $\psi_2$, succeeded by $\psi_4$, and so
forth, start gradually to increase.
On the contrary, the suppression of higher harmonics at low momenta
renders the ``spectral content'' of the DA $\varphi_{\pi}^{(2)}$ less
important.
The certain impact of importance and uncertainties of the different
contributions will be discussed in the next section.

\section{LCSR predictions for $F_{\gamma\pi}$ and their uncertainties}
\label{sec:LCSR-uncertainties}

In this section we identify the main sources of the various theoretical
uncertainties and estimate their effects on the computation of the
pion-photon TFF within the LCSR approach.

\subsection{Leading twist DA models and their uncertainties}
\label{subsec:DA-uncertainties}
\begin{figure}[bh]
\includegraphics[width=0.44\textwidth]{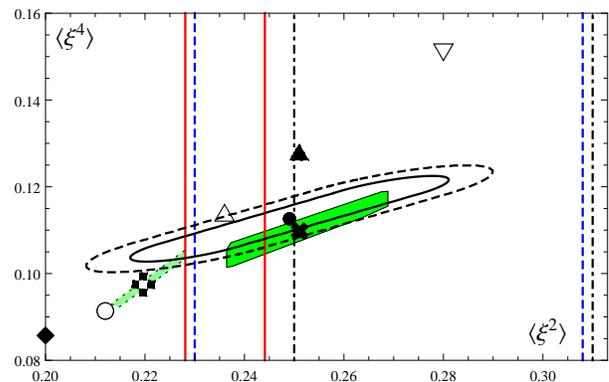} 
\caption{
Locations of various pion DAs projected onto the plane spanned by the
moments
$\langle \xi^2 \rangle$ and $\langle \xi^4 \rangle$
at the momentum scale $\mu=2$~GeV.
For those DAs which were originally determined at a lower normalization
scale, NLO evolution has been employed.
Upward pointing stretched green rectangle ---
BMS admissible region with $\lambda_{q}^{2}=0.4$~GeV$^2$,
including the BMS DA \ding{54};
\ding{60} --- platykurtic DA \cite{Stefanis:2014nla}
within the admissible region of similar DAs
determined in \cite{Stefanis:2015qha} with
$\lambda_{q}^{2}=0.45$~GeV$^2$;
$\bigcirc$ --- Light-Front DA \protect\cite{Choi:2014ifm};
\ding{117} --- asymptotic DA;
\ding{115} --- DSE-DB $\pi$ DA \protect\cite{Chang:2013pq};
$\bigtriangledown$ --- DSE-RL $\pi$ DA \protect\cite{Chang:2013pq};
$\bigtriangleup$ --- AdS/QCD $\pi$ DA \protect\cite{Brodsky:2011yv}.
The vertical lines denote the constraints extracted for
$\langle \xi^2 \rangle$
from various lattice simulations:
solid red lines --- \cite{Braun:2015axa};
dashed blue lines --- \cite{Braun:2006dg};
dashed-dotted lines --- \cite{Arthur:2010xf}.
1$\sigma$ (solid line) and 2$\sigma$ (dashed line) error ellipses
obtained with a LCSR-based fit to the CELLO \cite{CELLO91},
CLEO \cite{CLEO98}, BABAR($\leq$9~GeV$^2$) \cite{BaBar09},
and Belle \cite{Belle12} data.
\label{fig:BMSspace}
}
\end{figure}

The key nonperturbative input in the computation of the TFF is the pion
distribution amplitude of twist two, i.e.,
Eq.\ (\ref{eq:gegen-exp}), which depends on the conformal coefficients
$a_n(\mu^2)$.
In our approach $\{a_n(\mu^2)\}$ are obtained from QCD sum rules with
NLCs \cite{BMS01}, first proposed in \cite{MR86,MR89}.
The method in \cite{BMS01} allows us to extract at the typical hadronic
scale $\mu^2 \approx 1.35$~GeV$^2$ (emerging naturally in the approach)
a whole family of DAs.
This is done by fitting the sum rules for the first ten moments
\begin{equation}
  \langle \xi^{N} \rangle_{\pi}
\equiv
  \int_{0}^{1} dx (2x-1)^{N} \varphi_{\pi}^{(2)}(x,\mu^2) \, ,
\label{eq:moments}
\end{equation}
where $\xi = x - \bar{x}$, together with their uncertainties.
The DAs are then expressed in terms of a two-parametric model of
the generic form
\begin{equation}
   \varphi_{\pi}^\text{BMS}(x,\mu^2)
\!=\!
6x\bar{x}
 \left[
       1 + a_2 C_{2}^{(3/2)}(\xi) + a_4 C_{4}^{(3/2)}(\xi)
 \right].
\label{eq:truncated}
\end{equation}
This parametrization is defacto justified because all higher conformal
coefficients $a_n(\mu^2)$ were found by calculating the moments
$\langle \xi^{N} \rangle_{\pi}$ ($N=2,4,\ldots, 10$)
to be negligible but bearing a large margin for error, see
\cite{BMS01,Bakulev:2004mc} for details.
The admissible region of the first two moments
$\langle \xi^{2} \rangle_{\pi}$ and $\langle \xi^{4} \rangle_{\pi}$
is shown graphically in Fig.\ \ref{fig:BMSspace} in the form of an
upward pointing slanted (green) rectangle, with its center being
marked by the symbol \ding{54} and denoting the BMS DA \cite{BMS01}.
The associated pairs of $(a_2, a_4)$ values fit best all moments
$\langle \xi^N \rangle$ with $N=2,4,\ldots, 10$
within the estimated errors.
These moments were determined by employing the nonlocality parameter
$\lambda_{q}^{2}=0.4$~GeV$^2$.

One can compute the values of the moments and the conformal
coefficients at any desired momentum scale using the ERBL evolution
equation.
The symbol \ding{60} in this figure denotes the position of the
recently proposed \cite{Stefanis:2014nla} platykurtic (pk) pion DA,
obtained within the BMS approach but using the still admissible value
$\lambda_{q}^{2}=0.45$~GeV$^2$.
A whole region of such platykurtic DAs was determined subsequently
by two of us in \cite{Stefanis:2015qha} and is shown in this figure
in terms of the shorter rectangle in light-green color on the left
of the previous one.
It is worth mentioning that the platykurtic DA is a unimodal curve
with a flat peak at $x=1/2$ and suppressed tails at $x=0,1$.
The numerical values of the second and fourth moments of these
two sets of DAs have been calculated at the momentum scale $\mu=2$~GeV
after NLO evolution using the \MSbar scheme to obtain
\begin{eqnarray}
\!\! \text{\ding{54}}~\langle \xi^2 \rangle^\text{BMS}\!\!\!\!
& = & \!\!
  0.251_{-0.015}^{+0.018};~
\langle \xi^4 \rangle^\text{BMS}
 =
  0.110_{-0.008}^{+0.009};
\\
\!\! \text{\ding{60}}~\langle \xi^2 \rangle^\text{pk}
& = & \!\!
  0.220^{+0.009}_{-0.006};~
  \langle \xi^4 \rangle^\text{pk}
=
  0.098^{+0.008}_{-0.005} \, .
\label{eq:BMS-pk-moments}
\end{eqnarray}
The corresponding conformal coefficients
are given by
\begin{eqnarray}
a_{2}^\text{BMS}
& = & \!\!
 0.149_{-0.043}^{+0.052};~
a_{4}^\text{BMS}
 =
  -0.096_{-0.058}^{+0.063};
\\
a_{2}^\text{pk}
& = & \!\!
  0.057^{+0.024}_{-0.019};~
  a_{4}^\text{pk}
=
  -0.013^{+0.022}_{-0.019} \, .
\label{eq:BMS-pk-coefficients}
\end{eqnarray}
Note that the numerical values provided above for the moments of the
BMS DA are slightly different from those quoted in Table 1 of
\cite{Stefanis:2015qha}.
The reason for this discrepancy is that here we use a more advanced
code for the NLO evolution than that employed in \cite{Stefanis:2008zi}
and quoted in \cite{Stefanis:2015qha}.
The new code takes into account the quark-mass thresholds in more
accurate way and yields less suppression due to evolution.
We present these results at the momentum scale $\mu=2$~GeV
because this scale is commonly used in lattice calculations of
$\langle \xi^2 \rangle$ and $a_2$, as those indicated in the figure by
the vertical lines.
The solid red lines furthest to the left display the most recent
constraints determined
in \cite{Braun:2015axa},
while the dashed blue lines show the older results of the same group
\cite{Braun:2006dg},
and the dashed-dotted lines reproduce the regions of values computed in
\cite{Arthur:2010xf}.
The corresponding numerical values of the second moment, in the same
order of appearance, are
\begin{subequations}
\begin{eqnarray}
  \langle \xi^2 \rangle
& = &
  0.2361(41)(39) ~~\!\!\text{\cite{Braun:2015axa}}
\label{eq:lat15} \\
  \langle \xi^2 \rangle
& = &
  0.269(39) ~~~~~~~~\text{\cite{Braun:2006dg}}
\\
  \langle \xi^2 \rangle
& = &
  0.28(2)(1) ~~~~~~~\text{\cite{Arthur:2010xf}} \, .
\label{eq:lattice-moments}
\end{eqnarray}
 \end{subequations}
Note that the total error shown in Fig.\ \ref{fig:BMSspace} with
reference to Eq.\ (\ref{eq:lat15}) is the linear sum of the errors in
parentheses.
Assuming instead that these errors are statistically independent and
obey normal distributions, we would obtain by the sum in quadrature a
somewhat narrower range of constraints on $\langle \xi^2 \rangle$
than the vertical solid (red) lines.
A detailed treatment of the extraction of the conformal coefficients
from these lattice constraints is given in
\cite{Stefanis:2008zi,BMPS11,Stefanis:2015qha}.
On the other hand, the symbol $\bigcirc$ denotes the model DA from
\cite{Choi:2014ifm} extracted within a light-front-based framework.
This DA has a single broad peak and suppressed tails like the
platykurtic DA.

In Fig.\ \ref{fig:BMSspace} the asymptotic DA is also shown in terms
of the symbol \ding{117}, while \ding{115} and $\bigtriangledown$
represent, respectively, the DSE-DB and DSE-RL $\pi$ DAs
\cite{Chang:2013pq}, where the abbreviations are labels for the most
advanced kernel --- DB --- and the rainbow ladder (RL) approximation
in the use of Dyson-Schwinger equations (DSE) --- \cite{Raya:2015gva}.
In this figure, we also include the LCSR-based
(cf.\ Eq.\ (\ref{eq:LCSR-FQq}))
$1\sigma$ (solid black line) and
$2\sigma$ (dashed black line) error regions of the
CELLO \cite{CELLO91}, CLEO \cite{CLEO98}, and Belle \cite{Belle12}
data in terms of two parameters, viz., $\langle \xi^2 \rangle$ and
$\langle \xi^4 \rangle$.
The BABAR \cite{BaBar09} data below $Q^2 \leq 9$ GeV$^2$ have also been
taken into account.
One observes that there is a sizeable overlap between the BMS region of
bimodal DAs (larger green strip) and the data.
This overlap is also compatible with the lattice constraints.
The platykurtic region has a small overlap with the $1\sigma$ and
$2\sigma$ error ellipses, being at the same time just on the lower
boundaries of the lattice constraints on $\langle \xi^2\rangle$.
On the other hand, the broad, endpoint-enhanced DSE DAs (\ding{115} and
$\bigtriangledown$) conform within errors with the older lattice
constraints but disagree with the data up to the level of $2\sigma$.
It is fair to notice here that the authors of \cite{Raya:2015gva} argue
that their predictions for $Q^2F_{\gamma\pi}(Q^2)$, computed with a
QCD-based framework in terms of Dyson-Schwinger equations, agree with
the CELLO, CLEO, and Belle sets of data and thus belong to the green
band of predictions described in \cite{BMPS12}.
However, the truncation scheme in this approach cannot systematically
connect Eq.\ (\ref{eq:fact-TFF}) with the twist expansion.
The incompatibility between these DSE-based results and our findings
in Fig.\ \ref{fig:BMSspace}, obtained with a LCSR-based data fit,
demands further examination.

A similarly broad pion DA $(8/\pi)\sqrt{x\bar{x}}$, based on the
AdS/QCD and light-front holography, is displayed in this figure by the
symbol $\bigtriangleup$ \cite{Brodsky:2011yv}.
This DA appears to be just inside the upper boundary of the $2\sigma$
error ellipse of the experimental data.
The predictions for $Q^2F_{\gamma\pi}(Q^2)$ obtained with this pion DA
were found \cite{Brodsky:2011yv,Brodsky:2011xx} to agree well with the
CELLO and CLEO data, but to disagree with BABAR's large $Q^2$ data.
They belong to the green band of theoretical predictions in the
classification scheme of Ref.\ \cite{BMPS12} (see Fig.\ 2 there)
and conform with the Belle data as well.
As a final remark, we note that a faithful conformal expansion of such
broad DAs, like the DSE DAs and the holographic one, should include a
much larger number of terms of the order of 50.
Thus, the projection on the
($\langle \xi^2 \rangle$, $\langle \xi^4 \rangle$)
plane in Fig.\ \ref{fig:BMSspace} is a rather crude approximation for
such DAs, see \cite{Stefanis:2008zi,Stefanis:2015qha} for further
discussion.

\subsection{Higher-order radiative corrections}
\label{sec:hi-rad-cor}
In this subsection we discuss the uncertainties entailed by the NNLO
radiative corrections, entering the spectral density in
(\ref{eq:rho-n}).
To start with, recall Eq.\ (\ref{eq:NNLO-beta}) in conjunction with
the equations in (\ref{eq:hard-scat-nnlo-beta}) and (\ref{eq:T-elements}).
To continue, we reduce the full spectral density $\bar{\rho}^{(2)}$
to the expression
$
 a_s^{2}~\bar{\rho}^{(2)}_n(Q^2,x)
\to
 a_s^{2}~\beta_0 \bar{\rho}^{(2\beta)}_{n}(Q^2,x)
$,
cf. Eq.\ (\ref{eq:hard-scat-nnlo-beta}), ignoring this way all other
terms in Eq.\ (\ref{eq:NNLO-beta}).
This $\beta_0$ part of the spectral density is given in Appendix
\ref{sec:spectr-dens} and has already been used to obtain the NNLO
contribution to the TFF within the LCSR framework, see
\cite{MS09,Bakulev:2011iy,SBMP12,Stefanis:2014nla,Stefanis:2014yha}.
It turns out that this contribution is negative with a magnitude of
the order of 0.01~GeV to be compared with 0.1~GeV of the total
magnitude of the TFF at the generic hadronic boundary $Q^2=1$~GeV$^2$
of the pQCD applicability.
To increase the accuracy of the LCSR, we improve the treatment of the
NNLO contribution by taking into account in the spectral density
further terms related to expressions
(\ref{eq:hard-scat.nnlo-dv}) and (\ref{eq:hard-scat-nnlo})
of the hard-scattering amplitude.
This is a novelty of the present work and allows us fuller treatment
of the NNLO contribution and finer analysis of its uncertainties.

\begin{figure}[]
\includegraphics[width=0.45\textwidth]{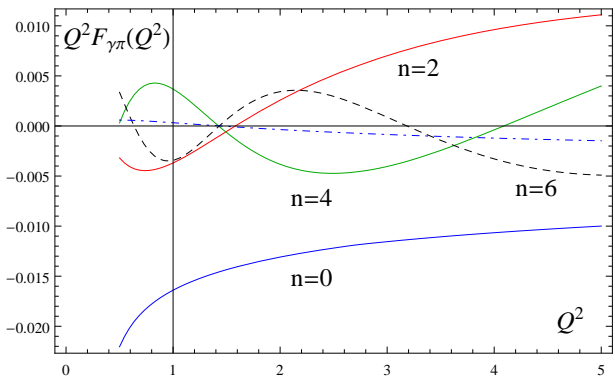} 
\caption{
Partial contributions $F^\text{NNLO}_n(Q^2)$ to the
TFF originating from the NNLO-$\beta_0$ term.
Only the results for the first Gegenbauer harmonics with $n=0,2,4,6$
are shown using the same notations as in Fig.\ \ref{fig:knot}.
The additional dashed-dotted (light-blue) flat line
represents the zero-harmonic contribution related
to the $\text{NNLO}-\Delta V$ term.
\label{fig:NNLO}}
\end{figure}

The uncertainty of the NNLO coefficient function
${\cal T}^{(2)}=\beta_0 {\cal T}^{(2)}_{\beta}+{\cal T}^{(2)}_{c}$
is induced by the yet uncalculated term ${\cal T}^{(2)}_{c}$,
whereas all other elements in $T_\text{NNLO}$ given by Eq.\
(\ref{eq:NNLO-beta}) are now known.
Lacking knowledge of the complete structure of the NNLO term, it seems
prudent to assume that the missing term ${\cal T}^{(2)}_{c}$ may have a
comparable magnitude as $\beta_0 {\cal T}^{(2)}_{\beta}$ which in turn
implies that the supposed uncertainty ensuing from our approximate
treatment will be rather overestimated.
In any case, each term in expression (\ref{eq:NNLO-beta}) for the
$T_\text{NNLO}$ hard-scattering amplitude entails an associated
contribution to the spectral density $\bar{\rho}^{(2)}_n$, notably,
\begin{eqnarray}
  \bar{\rho}^{(2)}_n
=
  C_{\rm F}\left(
  \beta_0\bar{\rho}^{(2\beta)}
  + \bar{\rho}^{(2\Delta V)}_{}
  + \bar{\rho}^{(2 L)}_{}
+{\cal T}^{(2)}_{c}
          \right)_n \, .
\end{eqnarray}
Here
$\bar{\rho}^{(2\beta)}$,
$\bar{\rho}^{(2\Delta V)}_{}$,
and
$\bar{\rho}^{(2 L)}_{}$
stem from Eqs.\ (\ref{eq:hard-scat-nnlo-beta}),
(\ref{eq:hard-scat.nnlo-dv}),
and (\ref{eq:hard-scat-nnlo}), respectively, so that
\begin{eqnarray}
  \bar{\rho}^{(2k)}_n
=
  C_{\rm F}^{-1} \text{Im}
  \left(
  T_0 \otimes T_k \otimes\psi_{n}
  \right) ~~~
  (k=\beta,\Delta V, L) \, , \label{eq:rho-T}
\end{eqnarray}
while the term ${\cal T}^{(2)}_{c}$ enters autonomously as in
(\ref{eq:NNLO-beta}).
According to our conjecture above, we will replace the unknown term
${\cal T}^{(2)}_{c}$ by
$\pm\beta_0{\cal T}^{(2)}_{\beta}$
inducing this way the discussed uncertainty
$\Delta\bar{\rho}^{(2)}_n=\pm\beta_0\bar{\rho}^{(2\beta)}$
in the spectral density
$\bar{\rho}^{(2)}_n$.
The final effect of these uncertainties on the TFF in the low-mid $Q^2$
region will be addressed later in Sec.\ \ref{sec:LCSR-predictions}.

To clarify the role of the partial NNLO radiative corrections, we
present in Fig. \ref{fig:NNLO} the NNLO-$\beta_0$ contribution to the
TFF, i.e., $F^\text{NNLO}_n$, for the first few terms up to $n=6$ of
the Gegenbauer-harmonics expansion in comparison with the
NNLO-$\Delta V$ contribution for the zero harmonic.
Taking into account that the NNLO-$L$ contribution is equal to zero for
the zero harmonic, we conclude from this figure that the additional
NNLO-$L$- and NNLO-$\Delta V$ terms can be safely ignored.

The main (negative in sign) contribution is provided by the
$\psi_0$-harmonic and is denoted by the lowest solid (blue) curve in
Fig.\ \ref{fig:NNLO}.
The higher harmonic contributions are smaller than this and oscillate.
Remarkably, they become positive but with a small delay in $Q^2$
relative to the LO case shown in Fig.\ \ref{fig:knot}.
Also the first knot is slightly shifted to the right and appears
at $\sim 1.4$~GeV$^2$.
The explicit expressions for the elements of $\bar{\rho}^{(1)}_n$ and
$\bar{\rho}^{(2)}_n$ are outlined in Appendix \ref{sec:spectr-dens}.

\bigskip
\section{Numerical results for $F_{\gamma\pi}$ in the low-$Q^2$
         spacelike domain}
\label{sec:LCSR-predictions}
\begin{widetext}
\begin{figure*}[th]
\includegraphics[width=0.60\textwidth]{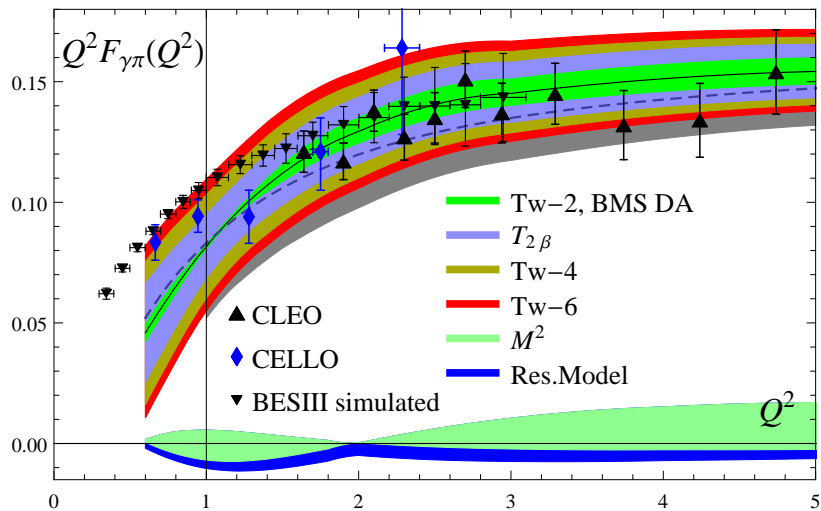} 
\vspace*{-3mm}
\caption{
Upgraded LCSR calculation of the pion-photon TFF using as
nonperturbative input the twist-two pion DAs obtained in
\protect{\cite{BMS01}} with $\lambda_{q}^{2}=0.40$~GeV$^2$
and taking into account NLO ERBL evolution.
The central wide (green) strip represents the result obtained with the
whole family of the BMS pion DAs, varying the conformal coefficients
$(a_2,a_4)$ within the appropriate region cf.\
(\ref{eq:BMS-pk-coefficients}) which corresponds to the slanted (green)
rectangle in the plane spanned by the associated moments
$
 \langle \xi^2 \rangle^\text{BMS},
 \langle \xi^4 \rangle^\text{BMS}
$
in Fig.\ \ref{fig:BMSspace}.
The central line inside the green strip shows the result for the BMS
model.
The uncertainties ensuing from different contributions are identified
in the graphics and are discussed in the text.
Taking into account the NNLO uncertainties gives rise to the violet
band next to the central green strip, whereas the variation of the
twist-four (Tw-4) and the twist-six (Tw-6) parameters generates (from
the inside to the outside) the orange and red outer strips,
respectively.
The two bands at the bottom of the figure show the additional
uncertainties originating from the variation of the Borel parameter in
the interval $M^2\in [0.7-1.5]$~GeV$^2$ (wide light-green band) and
the dependence on the modeling of the effective pion resonance in the
LCSR narrow blue strip), the latter being estimated as the difference
of the results obtained by using the BW model vs. the $\delta$-function
resonance model for the $\rho$ and $\omega$ resonances.
These two uncertainties have to be added to the ``rainbow'' band shown
on the top of the figure.
We also show in the graphics the influence of a non-vanishing small
virtuality of the quasireal photon in terms of the light-grey strip
below all others (see text for explanations).
The thick dashed line (close to the BMS solid line) corresponds to the
platykurtic model \cite{Stefanis:2014nla,Stefanis:2015qha} and serves
as a rough measure of the uncertainty induced by using
$\lambda_{q}^{2}=0.45$~GeV$^2$ inside the BMS approach.
The vertical line at 1~GeV$^2$ marks the typical applicability boundary
of our framework below which its reliability may become questionable.
\label{fig:errors}
}
\end{figure*}
\end{widetext}

Let us now discuss our LCSR-based calculation of the TFF in terms of
Fig.\ \ref{fig:errors} which effects graphically our core predictions
together with their various theoretical uncertainties worked out in
the previous section.
This analysis is bounded from below by the applicability limit of the
pQCD approach at the generic hadronic scale
1~GeV$^2$ which we indicated in this figure by a vertical line.
Although the obtained predictions are mathematically correct also
below this boundary, one cannot estimate their reliability from
the physical point of view.
Therefore, the displayed predictions below 1~GeV$^2$ only serve to
indicate the possible trend of the TFF in this momentum region.
The proper exploitation of the low-energy domain would demand
additional means, e.g., use of the axial anomaly exploited in
\cite{Klopot:2010ke,Klopot:2011qq,Klopot:2012hd} and
recently connected to the LCSRs in \cite{Oganesian:2015ucv}.

The considered uncertainties illustrated in Fig.\ \ref{fig:errors}
include
(i) the range of the admissible Gegenbauer conformal coefficients
$a_2$ and $a_4$ for the BMS DAs determined via QCD sum rules with
nonlocal condensates and using the nonlocality parameter
$\lambda_{q}^{2}=0.40(5)$~GeV$^2$ \cite{BMS01}
(narrow central green strip),
(ii) the result obtained by employing the platykurtic DAs derived with
the same method but with the slightly larger virtuality
$\lambda_{q}^{2}=0.45$~GeV$^2$ \cite{Stefanis:2014nla}
(thick dashed line slightly below the central strip),
(iii) the effect attributed to the unknown term ${\cal T}^{(2)}_{c}$ in
the NNLO contribution that has been approximated by
$\pm \beta_0{\cal T}^{(2)}_{\beta}$
(wide violet bands just above and below the central green strip),
(iv) the variation of the twist-four parameter
$\delta^2=0.19\pm 0.038$~GeV$^2$
in the range
$\delta^2\in[0.152-0.228]$~GeV$^2$
(light brown strips above and below the previous ones),
(v) the errors induced by the variation of the pre-factor
$(1^{+0.28}_{-0.23})\langle \sqrt{\alpha_s}\bar{q} q\rangle^2$
related to the uncertainty of the value of the quark condensate in
front of the twist-six expression in (\ref{eq:tw-6})
(red strips on the boundaries),
(vi) the effect of a small but finite virtuality of the quasireal
photon (strip in grey color below all the others) --- to be discussed
separately below,
(vii) the ambiguities in selecting the auxiliary Borel parameter
$M^2\in[0.7-1.5]$~GeV$^2$
(green band narrowing at 2~GeV$^2$ at the bottom),
(viii) the influence of the phenomenological description of the
resonance in the LCSR (narrow blue strip at the bottom) which displays
the difference between the results obtained from the Breit-Wigner
and the $\delta$-function resonance models.
These sources of systematic uncertainties have been collected for
convenience in Table \ref{tab:syserr} together with their partial
uncertainties ($\%$) at $Q^2=3$~GeV$^2$.

\begin{center}
\begin{table}
\caption{Sources and percentage estimates at $Q^2=3$~GeV$^2$ of the
systematic theoretical uncertainties in the LCSR-based calculation of
the pion-photon TFF illustrated in Fig.\ \ref{fig:errors}.
}
\begin{ruledtabular}
\begin{tabular}{lc}
Source & Uncertainty (\%) \\
       \hline \hline
Unknown NNLO term $\mathcal{T}_{c}^2$                & $\mp4.8$ \\
Range of Tw-2 BMS DAs                                & $-3.4 \div 4.1$ \\
Tw-4 coupling $\delta^2=[0.152-0.228]$~GeV$^2$       & $\pm 3.0$ \\
Tw-6 parameter variation                             & $-2.4 \div 3.0$ \\
Total                                                & $-13.6 \div 14.9$ \\ \hline
Borel parameter $M^2 \in [0.7-1.5]$~GeV$^2$          & $-1.6 \div 7.2$ \\
Resonance description $\delta$ vs. BW                & $-3.6 \div 0$ \\
Small virtuality of quasireal photon                 & $-5.4 \div 0$ \\
\end{tabular}
\end{ruledtabular}
\label{tab:syserr}
\end{table}
\end{center}
Focusing attention on the TFF in the vicinity of 1~GeV$^2$, we recall
our discussion of the correspondence of the two sides of
Eq.\ (\ref{eq:LCSR-pQCD}) to notice that in this momentum region mainly
the $\psi_0$-harmonic contributes, as illustrated in
Fig.\ \ref{fig:knot}.
This makes it evident that the contributions from different harmonics
in the vicinity of the knot at $Q^2 \sim 1$~GeV$^2$ vanish.

A possible small virtuality $q^2$ of the quasireal photon affects the
TFF and leads to an additional uncertainty of the predictions which
however is not universal but has to be estimated for each specific
experiment.
Theoretically, this effect can be expressed in terms of the
susceptibility $\Delta(Q^2)$ (linear response) which was invented in
\cite{SBMP12} (Sec.\ III there).
One has
\begin{eqnarray}
 \tilde{F}(Q^2, q^2)
& \approx &
  F(Q^2)
         \left[
               1 + \Delta(Q^2)  q^2
         \right] \ , \nonumber \\
  \Delta(Q^2)
&\equiv &
\frac{\tilde{F}'_{q^2}(Q^2,q^2=0)}{F(Q^2)} \, .
\label{eq:delta.Q2}
\end{eqnarray}

The susceptibility for the considered interval of $Q^2$ in
Fig.\ \ref{fig:errors} is approximately
$\Delta(Q^2) \simeq $ $-1$GeV$^{-2}$
as one can see from Fig.\ 3 in \cite{SBMP12}.
To get a qualitative estimate of this uncertainty and its influence on
the TFF, we use
$q^2\approx 0.04$~GeV$^2$,
which represents the maximal virtuality of the quasireal photon allowed
in the Belle experiment \cite{SBMP12}.
The result of the calculation is illustrated in Fig.\ \ref{fig:errors}
in terms of the lowest (grey) strip and has the tendency to reduce the
magnitude of the form factor in the whole range of $Q^2$ up to
asymptotic values, see \cite{SBMP12}.

Thus, from Fig.\ \ref{fig:errors} and Table \ref{tab:syserr} one may
conclude that for a given DA, the largest uncertainties in the
low-to-mid $Q^2\in [1-5]$~GeV$^2$ range originate from the NNLO
radiative correction and the twist-four and twist-six contributions.

\section{Conclusions}
\label{sec:concl}
The work presented here constitutes a systematic analysis of the
theoretical uncertainties entering the calculation of the pion-photon
transition form factor within the framework of LCSRs.
This method represents a very effective theoretical tool for the study
of this pion observable because it enables the sequential inclusion of
various contributions with controlled theoretical accuracy.
To be specific, we estimated the following main uncertainties:
(i) the relevance of the NNLO radiative corrections,
(ii) the ambiguity induced by the still unknown NNLO term
$\mathcal{T}^{(2)}_c $,
(iii) the influence of the twist-four and twist-six terms,
(iv) the sensitivity of the results on auxiliary parameters,
like the Borel scale $M^2$, and
(v) the role of the phenomenological description of resonances by
using a Breit-Wigner parametrization instead of a $\delta$-function
\textit{ansatz}.
Moreover, we computed the generic uncertainty pertaining to a small but
finite virtuality of the quasireal photon, albeit the precise magnitude
of this effect depends on the particular experimental setup.
A full list of the considered uncertainties and the estimation of their
size in percentage is given in Table \ref{tab:syserr} while a
visualization of these contributions to the scaled TFF is provided
in Fig.\ \ref{fig:errors}, focusing attention to the low-mid $Q^2$
region, where the BESIII Collaboration is expected to publish
high-statistics data in the near future.
The presented analysis complements and upgrades our previous works in
\cite{BMPS11,BMPS12,SBMP12}, in which our interest was primarily
concentrated on the high-$Q^2$ regime.
On the theoretical side, our study further extends the knowledge
of the NNLO contributions to the hard-scattering amplitude by
computing the terms $T_{\Delta V}$ and $T_L$ in
Eq.\ (\ref{eq:NNLO-beta}).
Moreover, we independently reproduced term-by-term all contributions
to the twist-six correction (\ref{eq:tw-6}), originally computed in
\cite{ABOP10}, and confirmed their validity.

\acknowledgments
We thank Nils Offen and Maksym Deliyergiyev  for useful discussions
and comments.
This work was partially supported by the Heisenberg--Landau
Program (Grants 2015 and 2016), the Russian Foundation for Basic
Research under Grants No.\ 14-01-00647 and No.\ 15-52-04023, and the
JINR-BelRFFR grant F16D-004.
A.V.P. was supported by the Chinese Academy of Sciences President's
International Fellowship Initiative (Grant No. 2016PM053),
the Major State Basic Research Development Program in China (Grant No.
2015CB856903), and the National Natural Science Foundation of
China (Grants No.\ 11575254 and No.\ 11175215).

\begin{appendix}
\appendix

\section{NLO evolution kernel and coefficient functions}
\label{sec:nlo-evol-kern}
In this appendix, the explicit expressions for the one- and two-loop
kernels of the ERBL evolution equation will be supplied, supplemented
by the coefficient functions.
We start by displaying the NLO evolution kernel
$
 V^{(1)}/C_{\rm F}
=
 \beta_0 V_{\beta}^{(1)} + \Delta V^{(1)}
$ in Eq.\ (\ref{eq:V-NLO}),
which has been computed in \cite{DR84,MR85}.
In order to reveal the origin of its individual contributions, we employ
the following new decomposition
\begin{widetext}
\begin{equation}
  V^{(1)}_{+} =
  C_{\rm F}
           \left\{
                  \Biggl[
                         \beta_0 V^{(1)}_{\beta +}
                        -C_{\rm F}\dot{V}^{(0)}_{+}\otimes V^{(0)}_{+}
                        -C_{\rm F}  \left[g_{+}, \otimes V^{(0)}_{+}\right]
                  \Biggr]
                + \left[
                        - 4\left(C_{\rm F}-\frac{C_{\rm A}}{2}\right)
                           \left(\frac{2}{3}V^{(0)}+2 V^{a}+ H\right)_{+}
                        +C_{\rm F} U_{+}
                  \right]
          \right\}
\label{eq:V1-structure}
\end{equation}
\end{widetext}
and discuss its structure term-by-term.
The first term, proportional to $\beta_0$, has the explicit form
\begin{equation}
  V^{(1)}_{\beta +}
=
  \left( \dot{V}^{(0)}+\frac{5}{3} V^{(0)}+2V^{a}\right)_{+}
\label{eq:Vbeta}
\end{equation}
and is related to the one-loop renormalization of $\alpha_s$
\cite{MR86ev,MS09}.
The second term
$-C_{\rm F}^{2} \dot{V}^{(0)}_{+}\otimes V^{(0)}_{+}$
results from the two-loop renormalization of the composite operator
\cite{MR86ev} and can be expressed as a convolution of one-loop
elements
\begin{widetext}
\begin{eqnarray}
&&
  \dot{V}^{(0)}_{+}\otimes V^{(0)}_{+}(x,y)
=
  2{\cal{C}}\theta(y>x)
                       \left\{\!\!\!\!\!\!\!\!\!\!\!\!\!\!\!\!\!\!\!\!\!\!\!\!\!\!\!\!\!\!\!\!\!\!\!
                               \phantom{\left(\frac{\bar X \ln\left(\bar X\right)}{Y}\right)}
  \left(F-\bar F\right)
                       \left[
                       \ln(y) \ln\left(\bar y\right)-\text{Li}_2(x)+\text{Li}_2(y)+\frac{\pi^2}{6}
                       \right]
\right. \nonumber \\ &&
  + \bar F
          \left[
    \text{Li}_2\left(1-\frac{x}{y}\right)-\text{Li}_2\left(1-\frac{\bar x}{\bar y}\right)
    +\ln\left(\bar x\right) \ln(xy)
    -\ln(y-x) \ln\left(\frac{\bar x}{\bar y}\right)
    -\frac{1}{2} \ln^2\left(\bar y\right)
          \right]
\nonumber \\ &&
  + F
    \left[
          \frac{3}{2}
          \ln\left(\frac{x}{y}\right)+\ln\left(\frac{x}{y}\right) \ln(y-x)-\frac{1}{2} \ln^2(x)
    \right]
          -\frac{11}{4}F+ 2 V^b
\nonumber \\ && \left.
          + \frac{x\bar x \left(\ln^2\left(\bar x\right)-2\ln(x) \ln(y)+\ln^2(y)\right)}{y\bar y (x-y)}
          -2 \left[
                   \frac{x \ln(y)}{\bar y}+\frac{\bar x \ln\left(\bar x\right)}{y}
             \right]
                      \right\} \, ,
\end{eqnarray}
\end{widetext}
where
$\displaystyle
 F(x,y)
=
 \frac{x}{y} \left(1+\frac{1}{y-x}\right)$
with
$\bar{F}=F(\bar{x},\bar{y})$.
Next we show the kernels $V^{(0)}$ and $\dot{V}^{(0)}$ in explicit form
\begin{widetext}
\begin{subequations}
\begin{eqnarray}
  V^{(0)}_+(x,y)
&=&
  2\left[{\cal C}\theta(y>x)\frac{x}{y}
         \left(1+\frac{1}{y-x}\right)
         \right]_{+} \equiv 2 \left[V^{a}(x,y) +V^{b}(x,y)\right]_{+}\, ,
\label{eq:V}
\\
  \dot{V}^{(0)}_{+}(x,y)
&=&
  2 \left[{\cal C} \theta(y>x)\frac{x}{y}
  \left(1+\frac{1}{y-x}\right)\ln\left(\frac{x}{y}\right)
  \right]_+ ,
\label{eq:Vdot}
\end{eqnarray}
where
\begin{eqnarray}
  V^{a}(x,y)
&=&
   {\cal C} \theta(y>x)\frac{x}{y},~~~~~V^{b}(x,y)
=
  {\cal C} \theta(y>x)\frac{x}{y}\left(\frac{1}{y-x}\right)\, ,
\label{eq:Def-va}
\end{eqnarray}
\end{subequations}
\end{widetext}
and the symbol $\mathcal{C}$ means
$ {\cal C}=\1+\left\{ x \to \bar{x}, y \to \bar{y}\right\} $.

Finally, the commutator
$\left[g_{+}, \otimes V^{(0)}_{+}\right]$ in (\ref{eq:V1-structure}),
which gives rise to the breaking of the conformal symmetry
\cite{Melic:2002ij,Belitsky:2000yn}, contains the element
\begin{widetext}
\begin{eqnarray}
  g_{+}(x,y)
&=&
  -2\left[
         \theta(y>x)\frac{\ln\left(1-x/y\right)}{y-x}
         +\theta(y<x)\frac{\ln(1-\bar{x}/\bar{y})}{x-y}
    \right]_{+} \, ,
\label{eq:g+}
\end{eqnarray}
so that with (\ref{eq:V}) we obtain
\begin{eqnarray}
&&
  \left[g_{+},\ \otimes V^{(0)}_{+}\right](x,y)
=
  -2{\cal{C}}\theta(y>x)
                        \left\{\!\!\!\!\!\!\!\!\!\!\!\!\!\!\!\!\!\!\!\!\!\!\!
                              \phantom{\frac{2(\ln(y))}{\bar y}}
   \left(F-\bar F\right)
                        \left(\text{Li}_2(y)-\text{Li}_2(x)
                        \right)
\right. \nonumber \\ &&
  + \bar F
          \left[
                \text{Li}_2\left(1-\frac{x}{y}\right)-\text{Li}_2\left(1-\frac{\bar x}{\bar y}\right)
                +\ln\left(1-\frac{x}{y}\right)\ln\left(\frac{x \bar y}{y \bar x}\right)
                +\frac{1}{2} \ln\left(\frac{\bar x}{\bar y}\right) \ln\left(\bar x \bar y\right)
          \right]
\nonumber \\ &&
  + \frac{1}{2} F
                 \left[
                       \ln\left(\frac{x}{y}\right)\ln\left(\frac{x y}{\bar x \bar y}\right)
                       -\ln(x y)\ln\left(\frac{\bar x}{\bar y}\right)
                 \right]
  -\frac{\pi ^2}{6} \left(F + \bar F \right)
\nonumber \\ && \left.
   -\frac{2}{y \bar y} \left(\bar x \ln\left(\bar x\right)-(y-x)
    \ln(y-x)+y \ln(y)\right)
\right\} \, .
\end{eqnarray}
\end{widetext}

\begin{widetext}
To complete the structure of the NLO evolution kernel $V_{+}^{(1)}$
entering Eq.\ (\ref{eq:V-NLO}), we also provide the expression for
$\Delta V^{(1)}$:
\begin{eqnarray}
  \Delta V^{(1)}_{+}
& = &
  \frac{1}{C_{\rm F}}V^{(1)}_{+} - \beta_0 V_{\beta+}^{(1)}
\nonumber \\
& = &
  -C_{\rm F} \dot{V}^{(0)}_{+}\otimes V^{(0)}_{+}
  -C_{\rm F} \left[g_{+}, \otimes V^{(0)}_{+}\right]
  - 4 \left(C_{\rm F}-\frac{C_{\rm A}}{2}\right)
      \left(\frac{2}{3}V^{(0)}+2 V^{a}+ H\right)_{+}
   +C_{\rm F} U_{+} \, .
\label{eq:A1c}
\end{eqnarray}
\end{widetext}

Note that the leading-order coefficient of the $\beta$ function used
in the above equations is
\begin{equation}
  \beta_0
=
  \frac{11}{3}{\rm C_A} - \frac{4}3 T_{\rm R} N_f\, ,
\label{eq: beta0}
\end{equation}
with $N_f$ being the number of active flavors ($N_f=4$ here) and
$T_{\rm R}=1/2, {\rm C_F}=4/3, {\rm C_A}=3$ for $SU(3)_c$.

The elements collected in the second square bracket in
(\ref{eq:V1-structure}) are all diagonal with respect to the one-loop
eigenfunctions $\psi_n$ by virtue of the symmetries
$U(x,y)y\bar{y}=x\bar{x}U(y,x)$ and $H(x,y)y\bar{y}=x\bar{x}H(y,x)$.
These quantities are displayed below for the convenience of the reader.
Note that the function $H(x,y)$ has been computed before, e.g.,
\cite{MR86ev}, while the function $U(x,y)$ was derived here.
\begin{widetext}
\begin{eqnarray}
  H(x,y)&=& {\cal C}\bigg\{   \theta(x> \bar{y})\left[
           2 (F - \bar{F})\text{Li}_2(1 - x/y) -
        2 F \ln(x) \ln(y) +
       (F-\bar{F})\ln^2(y) \right]
        \nonumber \\
      & & + 2 F\text{Li}_2(\bar{y}) \left[\theta(x> \bar{y}) - \theta(y > x)\right] +
        \theta(y > x) 2 \bar{F} \ln(y)  \ln(\bar{x}) \nonumber \\
      & &- 2 F\text{Li}_2(x) \left[\theta(x> \bar{y}) - \theta(x > y)\right]
             \bigg\} \, ,  \\
  U(x,y)&=&
  -\frac{5}{6}V^{(0)}+8 V^{a}- {\cal C} \theta(y>x) \bigg[4\frac{(y-x)}{y \bar{y}}\ln(y-x)\bigg]
\nonumber \\
&& +
  {\cal C} \theta(y>x) \bigg[2\frac{\bar x}{\bar y}\left(\frac{3y-1}{y}\ln(\bar{x}) + 2\ln(y)
      \right)-2\frac{x}{\bar y}\ln(y) \bigg]\, .
\label{eq:U}
\end{eqnarray}
\end{widetext}

\begin{widetext}
Finally, the coefficient functions of the partonic subprocess, described by
$\mathcal{T}^{(1)}$ and $\mathcal{T}^{(2)}_{\beta}$ in Eqs.\ (\ref{eq:T}),
(\ref{eq:hard-scat-nnlo-beta}) are
\begin{eqnarray}
   {\cal T}^{(1)}(x,y)
&=&
  \left[-3 V^{b}  +  g \right](x,y)_+ - 3 \delta(x-y),
\label{g} \\
 {\cal T}^{(2)}_{\beta}(x,y)
&=&
  \Bigg[\frac{29}{12} 2V^{a} + 2\dot{V}^{a}
        - \frac{209}{36} V^{(0)}  - \frac{7}{3} \dot{V}^{(0)}
        - \frac{1}{4} \ddot{V}^{(0)} + \frac{19}{6} g
        + \dot{g}
  \Bigg]_+\!\!(x, y)  - 6  \delta(x-y)\, .
\label{Def-C1Fnew}
\end{eqnarray}
\end{widetext}
The elements on the RHS of these equations were originally derived in
\cite{Melic:2002ij}, but are presented here in a different
notation following \cite{MS09}, where also the omitted elements
$\ddot{V}^{(0)}$ and $\dot{g}$ can be found.
\bigskip

\section{Elements of the spectral density $\bar{\rho}$}
\label{sec:spectr-dens}
Here we provide the contributions to the spectral density entering
Eq.\ (\ref{eq:rho-n}).
They are identified by the labels
$(0)$ --- LO term $\bar{\rho}^{(0)}_n$,
$(1)$ --- NLO term $\bar{\rho}^{(1)}_n$,
and $(2\ldots)$ NNLO terms, where the dots $\ldots$ indicate
particular contributions pertaining to the set of equations in (\ref{eq:T}).
For the default scale setting
$\mu_\text{R}^{2}=\mu_\text{F}^{2}=Q^{2}$,
they read
\begin{widetext}
\begin{eqnarray}
&&\bar{\rho}^{(0)}_n(x)=\psi_n(x) \, ,
\label{eq:barrho-0} \\ \nonumber
&& \bar{\rho}^{(1)}_n\left(Q^2=\mu^2_{\rm F};x\right)\frac{1}{C_{\rm F}}
=
    \left[ -3\left(1+v^{b}(n)\right)+\frac{\pi^2}{3}
           + 2v(n) \ln\left(\frac{\bar{x}}{x} \right)-\ln^2\left(\frac{\bar{x}}{x}\right)
           \right] \psi_n(x)  \\
&&    - 2 \left[  \sum^n_{l=0,2,\ldots}\!\!\!G_{nl}\psi_l(x)
                 +v(n)\left(\sum^n_{l=0,1,\ldots}\!\!\!b_{n l}\psi_l(x)-3\bar{x}\right)\right] \, , \label{eq:barrho-1-b}\\
&& v^{b}(n)= 2\left(\psi(2)-\psi(2+n) \right);~v(n)=1/(n+1)(n+2)-1/2 + 2\left(\psi(2)-\psi(2+n) \right) \label{eq:vb-v} \, .
\end{eqnarray}
\end{widetext}
The complete expression for $\bar{\rho}^{(1)}_n$ in
Eq.\ (\ref{eq:barrho-1-b}) was obtained in \cite{MS09}
and the content of the second square bracket was later corrected in
\cite{ABOP10} in the form it appears here.
The quantities $v^{b}(n)$ and $v(n)$ are the eigenvalues of the
elements $V^b_+$ and $V^a_++V^b_+$ of the one-loop kernel in
Eq.\ (\ref{eq:V}), respectively.
Expressions $G_{nl}$ and $b_{nl}$ denote the elements of calculable
triangular matrices (omitted here) --- see \cite{MS09,ABOP10}.
On the other hand, the $\beta_0$, $\Delta V$, and $L$ parts of the
NNLO spectral density have the following form
\begin{widetext}
\begin{eqnarray}
  \bar{\rho}^{(2\beta)}_{ n}\left(Q^2; x\right)
&=&
    {\cal T}_\beta^{(2)}(x,y)\otimes \psi_n(y)
    +
    \ln\left(\frac{\bar x}{x}\right)
    C_{1,n}(x)+C_{2,n}(x)-
\nonumber \\
&&
       v(n) \left\{
              \left[
                    \ln^2\left(\frac{\bar x}{ x}
              \right)
                     -\frac{\pi^2}3
        \right]
        \psi_n(x)
        +
        2\ln (x)\,C_{3,n}(x)-2C_{4,n}(x)
      \right\} \, , \label{eq:B4}\\
  \bar{\rho}^{(2\Delta V)}_{ n}\left(Q^2; x\right)
&=&
    \ln\left(\frac{\bar x}{ x}\right)
    \bar C_{1,n}(x) + \bar C_{2,n}(x) \, , \label{eq:B5}\\
 \!\!\!\! \bar{\rho}^{(2L)}_{ n}\left(Q^2; x\right)
&=&
    \ln\left(\frac{\bar x}{ x}\right)
    \tilde C_{1,n}(x) + \tilde C_{2,n}(x) + 2 C_{\rm F} v^2(n)
    \nonumber \\
&&
    \times \left\{
              \left[
                    \ln^2\left(\frac{x}{\bar x}
              \right)
                     -\frac{\pi^2}3
        \right]
        \psi_n(x)
        +
        2\ln (x)\,C_{3,n}(x)-2C_{4,n}(x)
      \right\} \, \label{eq:B6} \, ,
\end{eqnarray}
where we have introduced the auxiliary functions
\begin{subequations}
\label{eq:aux-f}
\begin{eqnarray}
  C_{1,n}(x) &=& (V^{(1)}_{\beta+}(x,y)-{\cal T}_1(x,y))\otimes\psi_n(y)\,,  \\
  C_{2,n}(x) &=& -\int\limits_0^{\bar x}\!\!du\, \frac{C_{1,n}(u)-C_{1,n}(\bar x)}{u-\bar x}\,, \\
  C_{3,n}(x) &=&  \int\limits_0^{\bar x}\!\!du\, \frac{ \psi_n(u)- \psi_n(\bar x)}{u-\bar x}\,, \\
  C_{4,n}(x) &=&  \int\limits_0^{\bar x}\!\!du\, \frac{ \psi_n(u)- \psi_n(\bar x)}{u-\bar x}\ln(\bar x-u) \, , \\
  \bar C_{1,n}(x) &=&\Delta V^{(1)}_+(x,y)\otimes\psi_n(y) \,,\\
  \bar C_{2,n}(x) &=& -\int\limits_0^{\bar x}\!\!du\, \frac{\bar C_{1,n}(u)-\bar C_{1,n}(\bar x)}{u-\bar x}\,, \\
  \tilde C_{1,n}(x) &=& 2 C_{\rm F} v(n) {\cal T}_1(x,y)\otimes\psi_n(y) \,,\\
  \tilde C_{2,n}(x) &=& -\int\limits_0^{\bar x}\!\!du\, \frac{\tilde C_{1,n}(u)-\tilde C_{1,n}(\bar x)}{u-\bar x}\, .
\end{eqnarray}
\end{subequations}
\end{widetext}

To derive the set of equations in (\ref{eq:aux-f}), we have used the
relations between the amplitudes, which contain powers of $L$,
and the various elements of the spectral density.
These relations are given by
\begin{eqnarray}
  \!\!\!\!\!\text{Im}\left[T_0\otimes \left(f L \right) \otimes \psi_n  \right]
\!\!&=&\!\!
  \ln\left(\frac{\bar x}{x}\right)C_{1,n}(f,x)+ \nonumber \\
  &&C_{2,n}(f,x)\,,  \label{eq:dens-fL} \\
  \!\!\!\!\text{Im}\left[T_0\otimes L^2 \otimes \psi_n  \right]
  \!\!&=&\!\!
	\left[
	\ln^2\left(\frac{\bar x}{x}
	\right)
	-\frac{\pi^2}3
	\right]
	\psi_n(x)+ \nonumber \\	
&&\!\!\!\!2\ln (x)\,C_{3,n}(x)-2C_{4,n}(x),
\label{eq:dens-LL}
\end{eqnarray}
where
\begin{eqnarray}
  C_{1,n}(f,x) \!\!&=& f(x,y)\otimes\psi_n(y)\,, \nonumber \\
  C_{2,n}(f,x) \!\! &=& \!\!\int\limits_0^{\bar x}\!\!du\, \frac{C_{1,n}(f,\bar x)-C_{1,n}(f,u)}{u-\bar x}\, .
\nonumber
\end{eqnarray}
\end{appendix}

\end{document}